\documentclass[useAMS,usenatbib]{mn2e}

\usepackage[pdftex]{graphicx}
\DeclareMathAlphabet{\mathpzc}{OT1}{pzc}{m}{it}

\newcommand\kms{\ensuremath{{\rm \, km\, s}^{-1} }}

\newcommand\varphis{\varphi_{\rm s}}
\newcommand\rs{r_{\rm s}}
\newcommand\zs{z_{\rm s}}

\title[Lensing by a singular isothermal sphere and a black hole]
      {Lensing by a singular isothermal sphere and a black hole}
\author[Mao \& Witt] {Shude Mao$^{1,2}$ and Hans J. Witt$^3$ \\
$^{1}$ National Astronomical Observatories, Chinese Academy of
Sciences, Beijing, 100012, China \\
$^{2}$ Jodrell Bank Centre for
Astrophysics, The University of Manchester, Alan Turing Building,
Manchester M13 9PL, UK \\
$^3$ Im Hollergrund 76, 28357 Bremen, Germany
}
\begin{document}
\include{journaldefs}
\date{Accepted ...... Received ...... ; in original form......   }

\pagerange{\pageref{firstpage}--\pageref{lastpage}} \pubyear{2011}
\maketitle
\label{firstpage}

\begin{abstract}
Most galaxies host central supermassive black holes. As two galaxies merge,
the black holes also merge. The final single black hole may suffer a kick due to asymmetric gravitational radiation and may not be at the centre of the galaxy; off-centre black holes may also be produced by
 other means such as sustained acceleration due to asymmetric jet power.
  We model the main galaxy as a singular isothermal sphere and the black hole as an off-centre point lens, and study the critical curves and caustics using complex notation. We identify the critical parameters that govern the transitions in the topology of critical curves, caustics and pseudo-caustics, and find the number of images can be two, three, four and five. We show examples of image configurations, including cases where three highly de-magnified images are found close to the centre. The perturbation on the image magnification due to the black hole scales linearly with its mass in the off-centre case, and quadratically when the black hole is at the centre. Such images are difficult to observe unless high-contrast and high-resolution imaging facilities (e.g., the Square Kilometer Array in the radio) become available.
\end{abstract}

\maketitle


\section{Introduction}
Most galaxies host central supermassive black holes (e.g., \citealt{bh2009} and references therein). As two galaxies merge,
the black holes at their centres may also merge. 
The two black holes' orbits first decay through dynamical friction when the
separation is large.  When the separation is very small, the black holes can merge efficiently through gravitational radiation. However, between these two limits, the black
holes may stall in their orbital decay. The stalling radius is typically at several pc to $\sim$ several tens of pc  (see, e.g., \citealt{2002MNRAS.331..935Y} and \citealt{2005LRR.....8....8M, 2009arXiv0906.4339C} for reviews). 

 Whatever brings the two black holes together (e.g., via gas processes), the final single black hole 
remnant may suffer a kick of the order of several thousand km/s due to asymmetric gravitational radiation, and thus may be off-centre (e.g., \citealt{pre07}). The black hole will oscillate at the centre of the galaxy, 
while its amplitude gradually declines on the timescale of $\sim 1$ Gyr (\citealt{gua08}).
An off-centre black hole can also be produced by sustained acceleration due to asymmetric jet power (\citealt{tsy07}).
Such a candidate has been reported in M87 (\citealt{m87-2010}) with an offset of approximately 12.8\,pc.	
	
Lensing by a single black hole at the centre of a singular isothermal ellipsoid has been studied by several authors (\citealt{mwk01, 2003ApJ...587L..55C, 2003A&A...397..415C, 2004ApJ...617...81B, 2005ApJ...627L..93R}).
 Motivated by observations as discussed above, we focus on off-centre black holes and study the critical curves and caustics using complex notation. For simplicity, we model the galaxy as a singular isothermal sphere.

The outline of the paper is as follows. In \S\ref{sec:center} we rederive the basic lensing results for a black hole at the centre. In \S\ref{sec:off-center} we study the case with an off-centre black hole, including the equations and topologies for critical curves and caustics, and then illustrate the image configurations. In \S\ref{sec:discussion} we discuss our results further in connection with observations. 

\section{Singular Isothermal Sphere plus a central black hole}
\label{sec:center}

In this section we investigate a model of a singular isothermal sphere (SIS)
plus a black hole at the centre. In complex notation (\citealt{wit90}), the lens equation is given by 
\begin{equation}
z_{\rm s}= z  - { m \over \bar{z}} - {z \over \sqrt{z \bar{z}}} 
= (r - {m \over r} - 1) e^{i\varphi},
\end{equation}
where $z=x+iy = r e^{i\varphi}$ and $z_{\rm s}= x_{\rm s}+ i y_{\rm s} = \rs e^{i\varphi_s}$ are the complex
coordinates of the lens and source plane respectively and ${\bar z}$ is the conjugate of $z$. The term with $m$ is from the black hole point lens while the square root term is from the SIS, here $m$ is the black
hole mass normalised by the total mass enclosed within the Einstein radius.

For a  lensing galaxy with velocity dispersion $\sigma=250\kms$ at redshift 0.5 and a source at redshift 2, the angular Einstein radius is $1.1$ arcsec in a cosmology with $\Omega_m=0.3$, $\Omega_\Lambda=0.7$ and Hubble constant $70\kms {\rm Mpc}^{-1}$. The enclosed (cylindrical) mass within the Einstein radius is $3.2 \times 10^{11} M_\odot$. From the correlation between the black hole mass ($M_{\rm bh}$) and $\sigma$ (\citealt{bh2009}), we find that $M_{\rm bh} \approx 2.4 \times 10^8 M_\odot$ which implies $m \approx 1.8 \times 10^{-3}$. The scatters in the black hole mass for a given $\sigma$ are quite large, 0.44\,dex in $\log M_{\rm bh}$. For definiteness we take $m=2.5 \times 10^{-3}$.
 
The above lens equation in polar coordinates can be easily transformed into a complex polynomial:
\begin{equation}
r^2 -r \rs e^{i(\varphi_s - \varphi)} - r - m = 0
\end{equation}
Due to the circular symmetry, all images must lie on a straight line.
Without loss of generality we assume all images are located on the $x$-axis
and  set $\varphi_s =0$ (i.e., the source is on the positive $x_{\rm s}$-axis). Since the solutions for $r$ must
be real and positive, $e^{i\varphi}$ must be restricted to $\pm 1$ for this
case. After some algebra, we find that there are always two images, given by
\begin{equation}
r_1 = {1 \over 2 } \left({\rs + 1} +  \sqrt{ (\rs + 1)^2+ 4m } \right),  
\quad \varphi = 0
\label{eq:phi=0}
\end{equation}
and
\begin{equation}
r_2 = {1 \over 2} \left({1-\rs} + \sqrt{ (1-\rs)^2+ 4m }\right),
\quad \varphi = \pi.
\label{eq:phi=pi}
\end{equation}

The (determinant) of the Jacobian of the lens mapping gives the inverse of the magnification:
\begin{equation}
J= \left( { \partial \zs\over \partial z } \right)^2 - 
 \left( { \partial \zs \over \partial \bar{z} } \right)
 \left( \overline{ { \partial \zs \over \partial \bar{z} }} \right)
 \label{eq:jacobian}
\end{equation}
with the derivatives
\begin{equation}
{ \partial \zs \over \partial z } = 1- {1 \over 2 \sqrt{z \bar{z} }} 
= 1 - {1 \over 2 r}
\end{equation}
and
\begin{equation}
{ \partial \zs \over \partial \bar{z} } = { m\over \bar{z}^2} +  
{ \sqrt{z \bar{z} } \over 2 \bar{z}^2}
= \left( {m\over r^2 } + { 1 \over 2 r} \right) e^{2i\varphi}.
\end{equation}
The critical curve is given by $J=0$. For our case, in polar coordinates, this condition is given by
\begin{equation}
(m + r^2) (m + r -r^2)  = 0.
\end{equation}
The first term is always positive while the second term is quadratic in $r$,
which yields one real positive solution 
(the negative solution is unphysical)
\begin{equation}
r_{c.c.} = {1\over 2} + \sqrt{ { 1 \over 4 } + m}.
\end{equation}
This critical curve is thus a ring with radius given by the above equation. It maps
into a degenerate caustic point at the origin, as expected from the
axis-symmetry. 
For completeness we derive also the magnification of the images (including parity)
which is given by
\begin{equation}
\mu_{1} = {1 + \rs \over 2 \rs} + 
{ (1+ \rs)^2 + 2m \over 2\rs \sqrt{(1 + \rs)^2 + 4m} } ,
\end{equation}
\begin{equation}
\mu_{2} =- {1 - \rs \over 2 \rs} -
{ (1- \rs)^2 + 2m \over 2\rs \sqrt{(1- \rs)^2 + 4m} }.
\end{equation}
For $m=0$, we recover the familiar magnifications for a SIS: $\mu_1=(1+\rs)/\rs$ and
$\mu_2=-(1-\rs)/\rs$ for $\rs \le 1$. In astrophysical applications, we expect $m \ll 1$, so we Taylor expand the above expressions into series of $m$ for $\rs < 1$:
\begin{equation}
\mu_1 \approx {1 + \rs \over \rs}+\frac{m^2}{(1+\rs)^3 \rs},
\mu_2 \approx -{1-\rs \over \rs} - \frac{m^2}{(1-\rs)^3 \rs}.
\label{eq:approx1}
\end{equation}
The perturbations on the magnifications scale as $m^2$. 

For  $\varphis=0$ and $\rs > 1$, the black hole creates a new image which does not exist if we have the SIS alone. The approximation for $\mu_1$ is still valid, but that for the magnification of the new image ($\mu_2$) has to be changed to
\begin{equation}
\mu_2 \approx - \frac{m^2}{(\rs-1)^3 \rs}.
\label{eq:approx2}
\end{equation}
Both approximations (eqs. \ref{eq:approx1} and \ref{eq:approx2}) for $\mu_2$ break down when $\rs \rightarrow 1$. 

\section{SIS with an off-centre black hole}
\label{sec:off-center}

We now consider the case with a SIS and an off-centre black hole. The lens equation is given by 
\begin{equation}
\zs = z  - { m \over \bar{z} - \bar{z}_0 } -  \frac{z}{\sqrt{z \bar{z}}},
\label{eq:lens}
\end{equation}
where $z_0$ denotes the position of the off-centre black hole in complex notation. We choose the coordinate system such that the black hole is on the positive $x$ axis with $z_0=r_0>0$.

\subsection{Images}
\label{sec:imageNumber}

To solve the lens equation for the image positions it is better to switch
from the Cartesian coordinates to polar coordinates. We again write
$z = r e^{i \varphi}$ and $\zs = \rs e^{i\varphis}$ ($r \ge 0, \rs \ge0$). 
The lens equation can now be written as
\begin{equation}
\rs e^{i\varphis} = r e^{i \varphi}  - { m \over  r e^{-i\varphi} - r_0} -  e^{i\varphi} .
\end{equation}
It is interesting to note that this equation differs only
by a missing $r$ in the last term compared with the binary lens equation
(assuming one point mass is located at the origin). 

If we clear now the fractions of the equation and 
take the real and imaginary parts of the equation one obtains two equations:
%
\begin{eqnarray}
r_0 \rs \sin\varphis + r \rs \sin(\varphi-\varphis) &=& r_0 r \sin\varphi - r_0
\sin\varphi \label{imageeq1}, \\
r_0 \rs \cos\varphis - r \rs \cos (\varphi - \varphis) &=& r_0 r \cos\varphi -
r_0\cos\varphi \nonumber \\
& & - r^2 + r + m.   \label{imageeq2}
\end{eqnarray}
The first equation can be manipulated to yield an expression for $r$:
\begin{equation}
r  = { r_0 \rs \sin\varphis + r_0 \sin\varphi \over r_0 \sin\varphi - \rs
  \sin(\varphi-\varphis) }.
\end{equation}
We may insert this equation into the second equation and obtain
an equation solely in $\varphi$.
This equation is equivalent to a polynomial of degree 6 in $e^{i\varphi}$
and can only be solved numerically in general, and so the maximum number of images is 6. In practice, we find the number of images can range from 2 to 5. For concrete examples, see \S\ref{sec:imageNumber}.

For the case when the source position is located on the $x_{\rm s}$-axis ($\varphis = 0, \pi$) the
equation still factorizes. When the source is on the positive $x_{\rm s}$ axis ($\varphis=0$), eqs. (\ref{imageeq1}) and (\ref{imageeq2})
simplify to
\begin{eqnarray}
\varphi = 0 \quad {\rm and}\quad r^2-r(\rs+r_0+1) -m+r_0 + \rs r_0 = 0, \label {eq:varphi1} \\
\varphi = \pi \quad {\rm and} \quad r^2+r(\rs+r_0-1) -m-r_0 + \rs r_0 = 0, \label{eq:varphi2}
\end{eqnarray}
and
\begin{eqnarray}
r &=& {r_0 \over r_0 - \rs} \quad {\rm and}  \nonumber \\ 
\cos \varphi &=& {(r_0 \rs - m) (r_0 - \rs) \over 2 r_0 \rs} + { 1 \over 2 (r_0 - \rs )}
\leq 1.
\label{eq:cosPhi}
\end{eqnarray}
Each equation can have two solutions. However, the latter one
is bound by the amplitude of $\cos\varphi$ and $r>0$.
In Appendix \ref{sec:maximum}, we show in this case, the achievable maximum number of images is five. Intuitively, this can be understood as follows -- there are either zero or two off-axis solutions due to symmetry with respect to the $x$-axis; on the $x$-axis, there are at maximum three solutions: two solutions where the black hole and SIS have deflection angles with the same sign, and one solution where the black hole and SIS have opposite signs.  Similar equations can be found for the case when the source is on the negative $x$-axis ($\varphis=\pi$), but are not presented here.

It is difficult to derive analytically the magnification
of the images for the general case. However, we can do so when the source is located on the (positive) $x_{\rm s}$-axis ($\varphi_s =0$) because as we mentioned before the lens equations factorizes (see eqs. \ref{eq:varphi1} to \ref{eq:cosPhi}).
Appendix \ref{sec:magOnAxis} gives more details. We mention here that the perturbation on the magnification is linear with respect to $m$, in contrast to the $m^2$ scaling when the black hole is at the centre.

\subsection{Critical Curves and Caustics}

The critical curves are given by the determinant
of the Jacobian $J = 0$ where $J$ is given by eq. (\ref{eq:jacobian}),
with the derivatives
\begin{equation}
{ \partial \zs \over \partial z } = 1- {1 \over 2 \sqrt{z \bar{z} }}
\label{eq:zsz}
\end{equation}
and
\begin{equation}
{ \partial \zs \over \partial \bar{z} } = { m\over (\bar{z_0} - \bar{z})^2} +  
{ \sqrt{z \bar{z} } \over 2 \bar{z}^2}.
\end{equation}
In polar coordinates, the determinant of the Jacobian is given by
\begin{equation}
J=1-\frac{1}{r} - \frac{m}{r} \,\frac{r^2 - 2 r r_0 \cos\varphi + r_0^2 \cos 2\varphi}{w^2}-\frac{m^2}{w^2},
\label{eq:J}
\end{equation}
where $w=r^2 - 2 r_0 r \cos\varphi + r_0^2, r>0, r_0>0$.

The critical curves are given by $J=0$, which can always be transformed from a
two-dimensional problem into a one-dimensional one using the parametric representation (see eq. 8 in \citealt{wit90}), resulting in
the following form
\begin{equation}
 { \partial \zs \over \partial z } e^{i\alpha}  = 
  { \partial \zs \over \partial \bar{z} } 
 \quad \mbox{with} \quad 0\le \alpha < 2\pi,
\end{equation}
since the first derivative in eq. (\ref{eq:zsz}) is always real. 
The equation needs to be solved in $z$ for each $\alpha$ in the range $0$ to $2\pi$.
If we switch to polar coordinates we can write the previous equation as
\begin{equation}
 ( 1 - {1 \over 2 r}) e^{i\alpha}  = 
  { m \over (r_0 - r e^{-i\varphi})^2 } + { e^{2i\varphi} \over 2 r}.
  \label{para}
\end{equation}
If we clear this equation of fractions and take the real
and imaginary parts of the equation one obtains two equations
parameterised by $\alpha$, which give
the coordinates in $r,\varphi$ of the critical curve:
\begin{eqnarray}
r_0^2 \cos(4\varphi - \alpha) - 2r r_0 \cos(3\varphi -\alpha) \nonumber \\
+ (2mr + r^2) \cos(2\varphi - \alpha)  \nonumber \\ 
+ (1-2r) (r_0^2 \cos(2\varphi) -2r r_0 \cos\varphi + r^2) = 0,  \label{eq:para1} \\
r_0^2 \sin(4\varphi - \alpha) - 2r r_0 \sin(3\varphi -\alpha) \nonumber \\
+ (2mr + r^2) \sin(2\varphi - \alpha)  \nonumber \\ 
+ (1-2r) (r_0^2 \sin(2\varphi) -2r r_0 \sin\varphi ) = 0. \label{eq:para2}  
\end{eqnarray}
For the starting point $\alpha = 0$ we can obtain
analytical results for on-axis solutions: 
\begin{equation}
\varphi = 0, \pi \quad \mbox{and} \quad mr + (1-r)(r_0 \mp r)^2 = 0. \label{start}
\end{equation}
This equation yields one to three solutions on the $x$-axis.
However, eqs. (\ref{eq:para1}) and  (\ref{eq:para2}) can also have off-axis
solutions which are more difficult to obtain. Therefore we need to disentangle
both equations and derive one equation solely in $r$ or $\cos \varphi$.
These equations (eqs. \ref{eq:start_r} and \ref{eq:start_phi}) are presented in Appendix \ref{sec:parastart}.
Using these starting points one may solve the whole critical curve numerically
by increasing the parameter $\alpha$ from $0$ to $2\pi$.

\begin{figure}
\includegraphics[scale=0.40, bb=0 0 385 567]{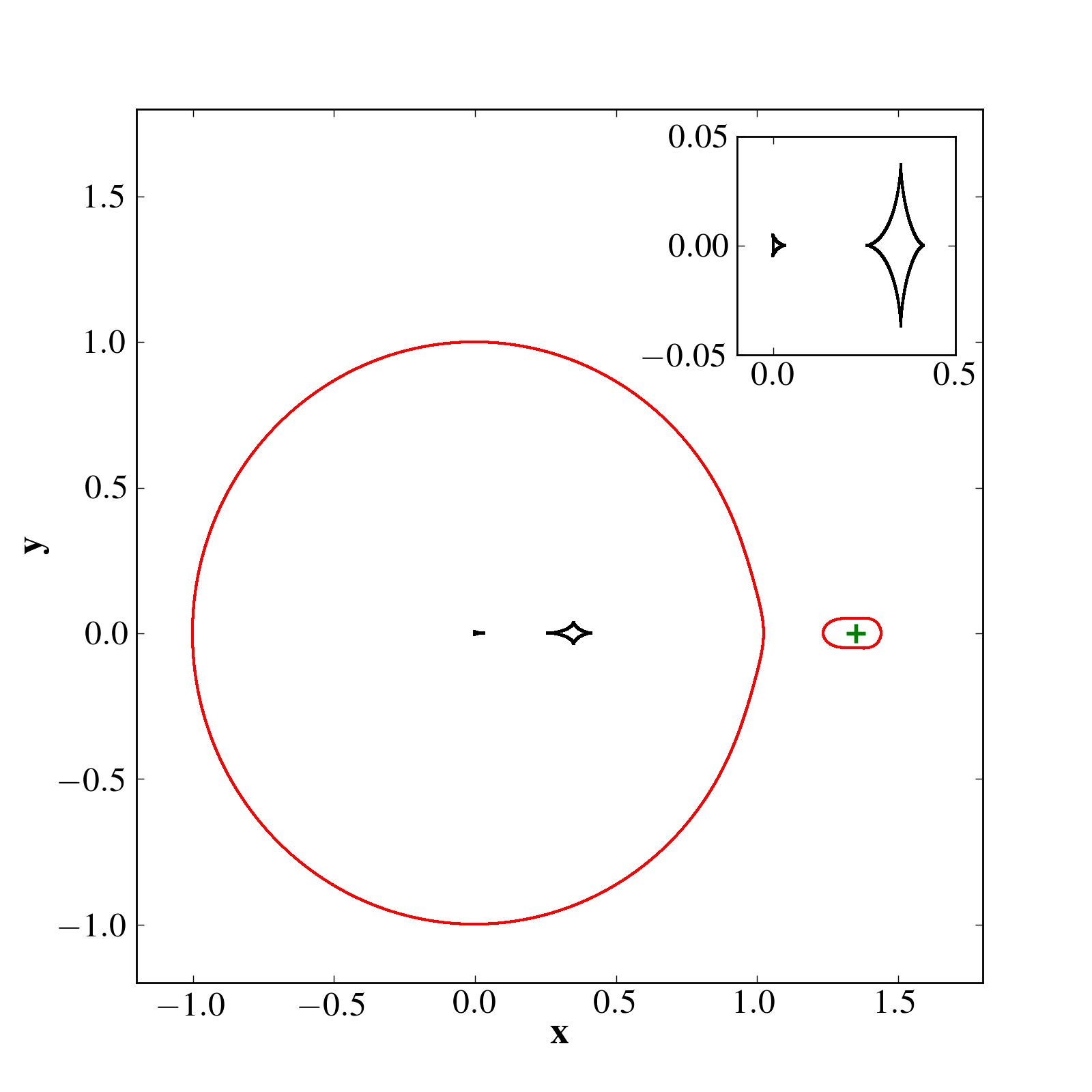}
\caption{critical curves (red) and caustics (black) for a singular
  isothermal sphere (SIS) plus a black hole. The SIS is centred at the
  origin. The black hole is at a distance of $r_0=1.35$ (indicated by a cross). A magnified view of the central region is shown on the top right. 
The left caustic (if further magnified) is similar to the one shown in the inset in Fig. \ref{fig:transition3}.
\label{fig:critical_1_35}
}
\end{figure}

\begin{figure}
\includegraphics[scale=0.40,bb=0 0 385 567]{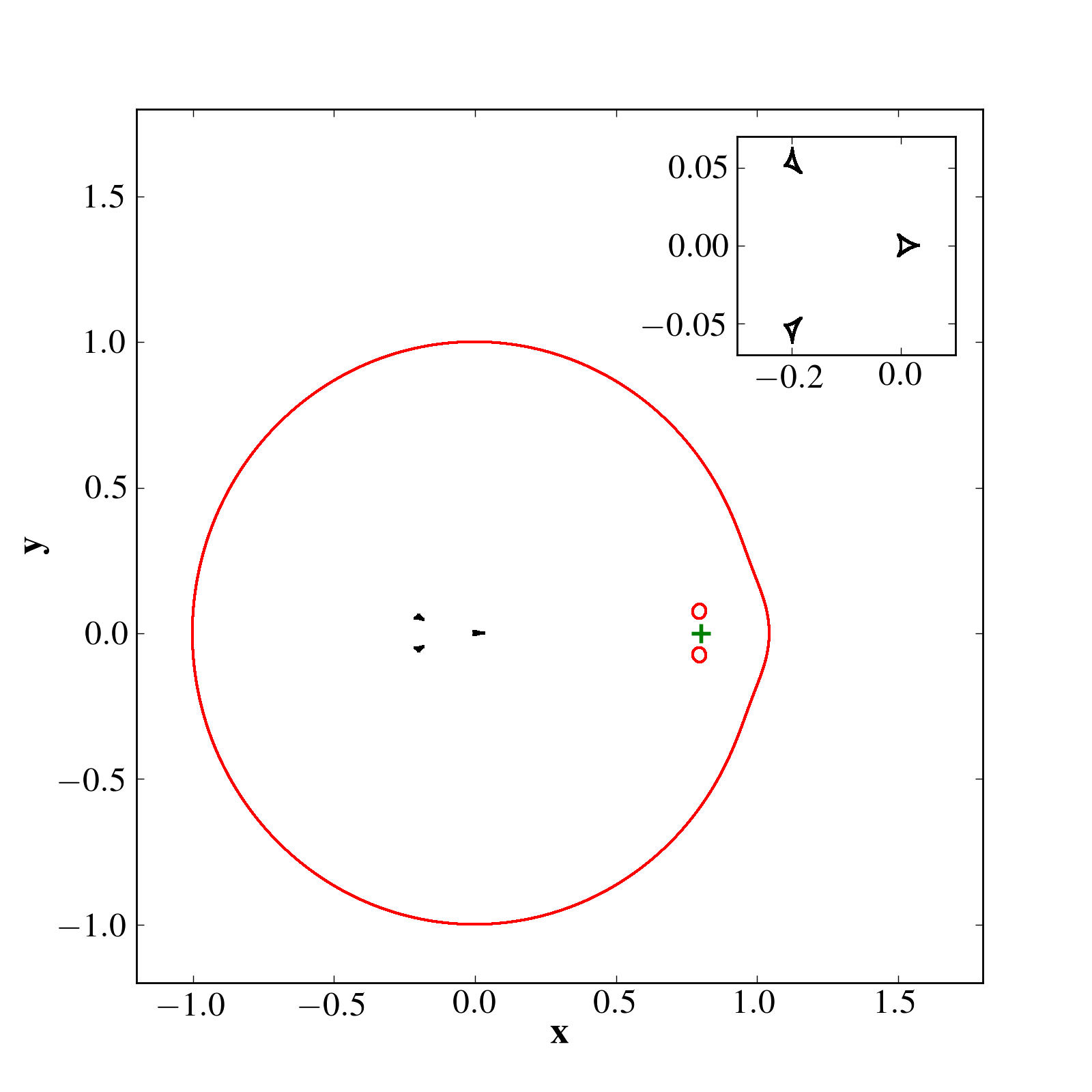}
\caption{Same as Fig. \ref{fig:critical_1_35} except $r_0=0.8$. The critical curves and caustics both have three segments.}
\label{fig:critical_0_8}
\end{figure}

\begin{figure}
\includegraphics[scale=0.40,bb=0 0 385 567]{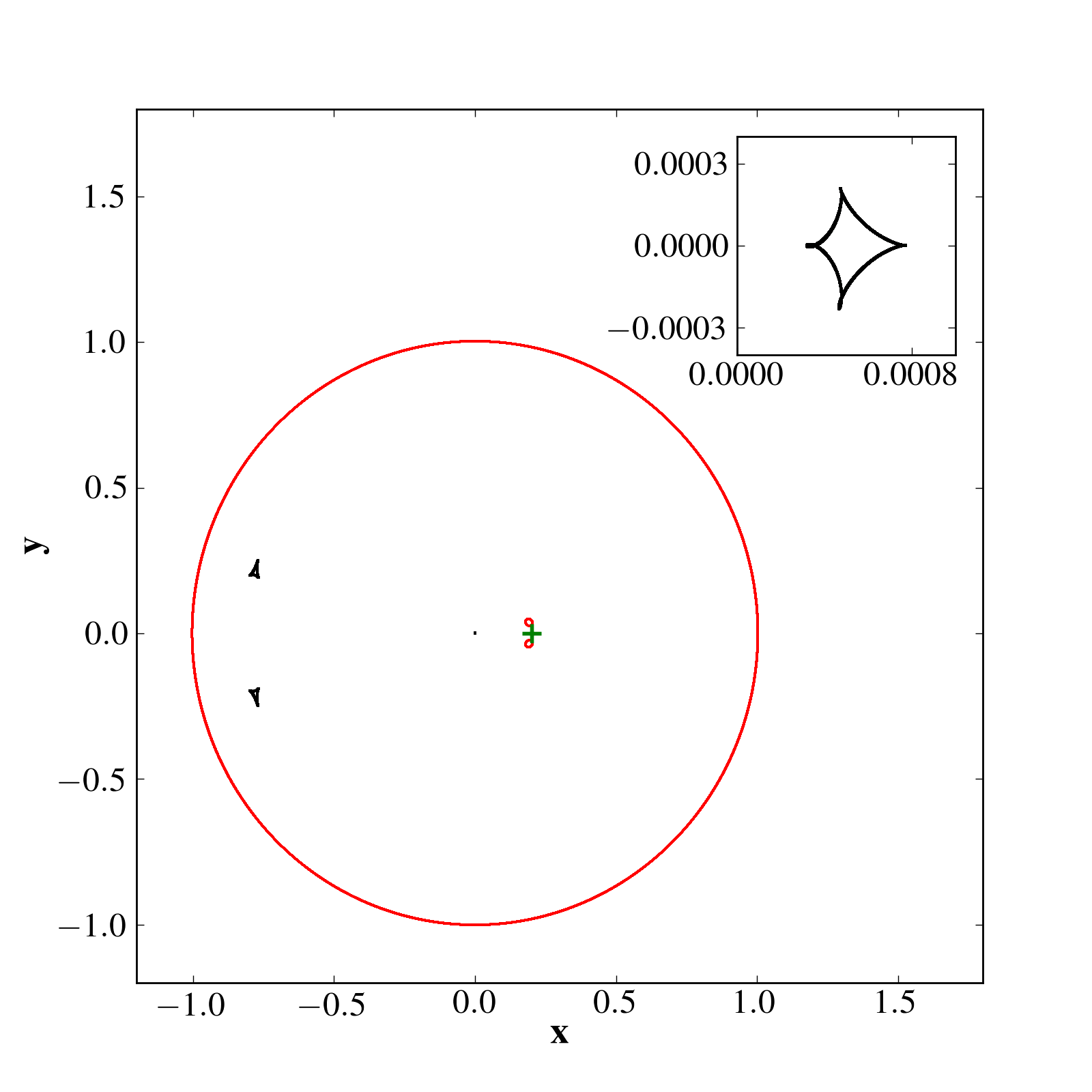}
\caption{Same as Fig. \ref{fig:critical_1_35} except $r_0=0.2$. }
\label{fig:critical_0_2}
\end{figure}

A transition in the topology of critical curves can take place
if the following conditions hold (\citealt{es93}):
\begin{equation}
J (r,\varphi) = 0, \quad {\partial J \over \partial r} = 0,
\quad {\rm and} \quad {\partial  J \over \partial \varphi} = 0.
\end{equation}
We can use the resultant method (\citealt{es93}) to
eliminate  $r$ and $e^{i\varphi}$ which yields the condition
for the transitions of the critical curves. Appendix \ref{sec:appendix} gives the technical details. For $m=2.5\times 10^{-3}$, we have 4 transitions at $r_0$ equals
\begin{enumerate}
\item
$r_{t1}=1.26361.$ 
\item $r_{t2}=0.891231$. 
\item $r_{t3}=\sqrt{1/4+m}$.
\item $r_{t4}=\sqrt{m}$.
\end{enumerate}
We will illustrate these transitions by examining the shapes of critical curves and caustics as we gradually decrease $r_0$ in a series of figures. As can be seen from Fig. \ref{fig:critical_1_35}, when $r_0>r_{t1}$ there are two disjoint critical curves. One is approximately a unit circle associated with the SIS, and the other is a small Einstein ring associated with the point lens. These two critical curves are mapped into two diamond caustics (see the inset). At $r_0=r_{t1}$ the two critical curves merge into a single one (see Fig. \ref{fig:transition1}), and remains so for $r_{t1}>r>r_{t2}$. For $r_0=r_{t2}$, the critical curve starts to split into three (see Figs. \ref{fig:transition2} and \ref{fig:critical_0_8}), with a primary critical curve associated with the SIS enclosing two ``holes". As the separation further decreases, the two ``holes" vanish, leaving behind only a single critical curve at $r_0=r_{t3}$ (Fig. \ref{fig:transition3}). At even smaller separations, the critical curves again split into three separate curves (see Fig. \ref{fig:critical_0_2}). Another `peculiar' transition occurs when $r_0=\sqrt{m}$ where the origin becomes part of the caustic (see Figs. \ref{fig:transition4}-\ref{fig:images}). This will be discussed in more detail below.

\subsubsection{Critical curves going through $r=0$}
\label{sec:r=0}

For a SIS plus a black hole we may have a special
transition when the critical curve is attached to the origin.
In this case we have $r=0$ and the polar coordinate $\varphi$ does not need
to have a particular value. We investigate eq. (\ref{para})
for the condition for this to occur. 

To do this, we first clear the equation of fractions
and obtain a polynomial of degree 4 in $e^{i\varphi}$.
We may now take the complex conjugate of the equation (exchanging $\varphi$ by
$-\varphi$ and $\alpha$ by $-\alpha$) and multiply the new equation by
$e^{4i\varphi}$. We thus obtain two linear independent equations where we
can eliminate $e^{i\varphi}$ using the resultant method. Performing these steps, we obtain
a  polynomial in $r$ of the form $4 e^{i2\alpha} r^2 P_{22} (r) = 0,$ 
where $P_{22}(r)$ is a polynominal of degree 22 in $r$ (which is too cumbersome to present here). The constant term of this polynomial is given in the left hand side of eq. (\ref{minalpha}).

We note here that $r=0$ can not be a generic solution of eq. (\ref{para}). The polynomial $p_{22}(r)$ has a non-trivial
solution for $r=0$ if the constant term in $P_{22}$ vanishes, i.e., 
\begin{equation}
m e^{i2\alpha} - 2 r_0^2 e^{i\alpha} + m =0. \label{minalpha}
\end{equation}
It is interesting to note that this equation has a valid solution for $\alpha$
if $r_0^2 \leq m $. This means that if the singularity of the SIS is inside the
Einstein ring of the black hole the critical curves are attached to the
origin.

In particular, for the case $m = r_0^2$ the two inner critical curves start to attach
to the origin, and is one of the transition points described above. In this case we have $r=0$, $\alpha = 0$ and 
$\varphi = \pm \pi / 2$.
For $r_0 < \sqrt{m}$, $\alpha =0$ may no longer be a solution
for the inner critical curves. However, eq. (\ref{minalpha}) defines the
minimum $\alpha$ value which yields a solution (and the starting point) at the origin.  
That means that the pseudo caustic (see the next subsection) becomes part of the
caustics (see Fig. \ref{fig:critical_0_025}). This is the first case we are aware of
in the literature where  a pseudo caustic merges with a real caustic.
Furthermore the singularity of the isothermal sphere 
starts to swallow solutions for the parametric
representation $\alpha$. Note that for $r_0\rightarrow 0$ the starting point
for $\alpha$ moves to $\pm \pi/2$.

 \subsubsection{Pseudo-caustic}

\begin{figure}
\includegraphics[scale=0.40,bb=0 0 385 567]{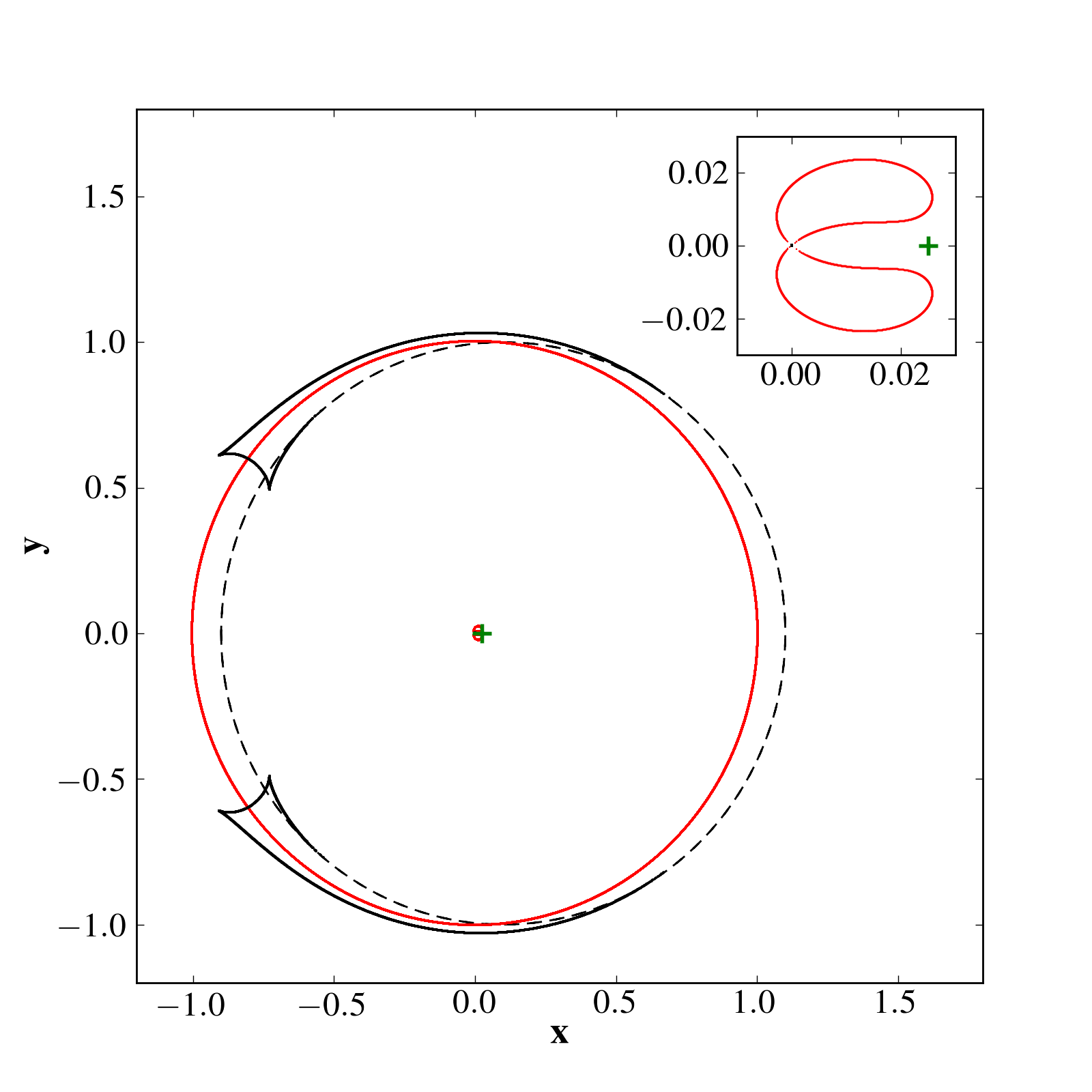}
\caption{
critical curves (red) and caustics (black) for a singular
  isothermal sphere (SIS) plus a black hole. The SIS is centred at the
  origin. The black hole is at a distance of $r_0=0.025$ (indicated by a cross). A magnified view of the central region is shown on the top right. The dashed black curve shows the pseudo-caustic corresponding
to $r = 0$, which gives
$\zs=m/r_0-\exp(i \varphi)$ where $0 \le \varphi < 2 \pi$. 
}
\label{fig:critical_0_025}
\end{figure}

When a source crosses a true caustic, the image number changes by two. In contrast, when a source crosses a pseudo-caustic, the image number changes by one (\citealt{ew98}). The deflection angle for the SIS is not continuous at $r=0$ due to the singularity, which gives rise to a pseudo-caustic. Since $z=r e^{i \varphi}$, as $r \rightarrow 0$, the lens equation (\ref{eq:lens}) maps into the source plane as
\begin{equation}
\zs=\frac{m}{r_0}-e^{i \varphi},~~ 0 \le \varphi < 2 \pi.
\end{equation}
This is a unit circle with the origin at $(m/r_0, 0)$, shown as the dashed black curve in Fig. 
\ref{fig:critical_0_025}. 

\begin{figure}
\includegraphics[scale=0.40,bb=0 0 385 567]{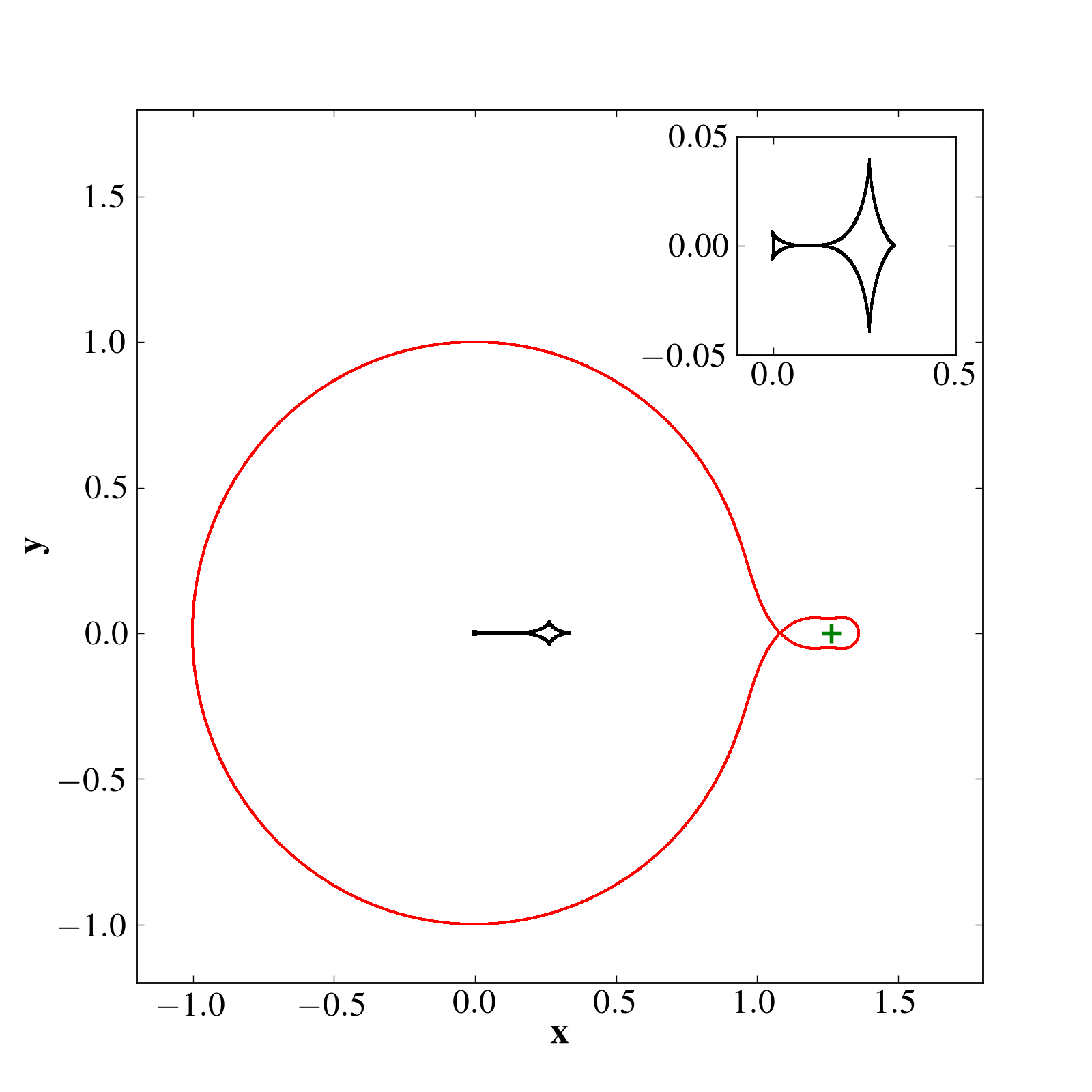}
\caption{Same as Fig. \ref{fig:critical_1_35} except $r_0=r_{t1}=1.26361$. Notice that
  the two isolated critical curves have merged into a single one, so have
  the caustics.
}
\label{fig:transition1}
\end{figure}

\begin{figure}
\includegraphics[scale=0.40,bb=0 0 385 567]{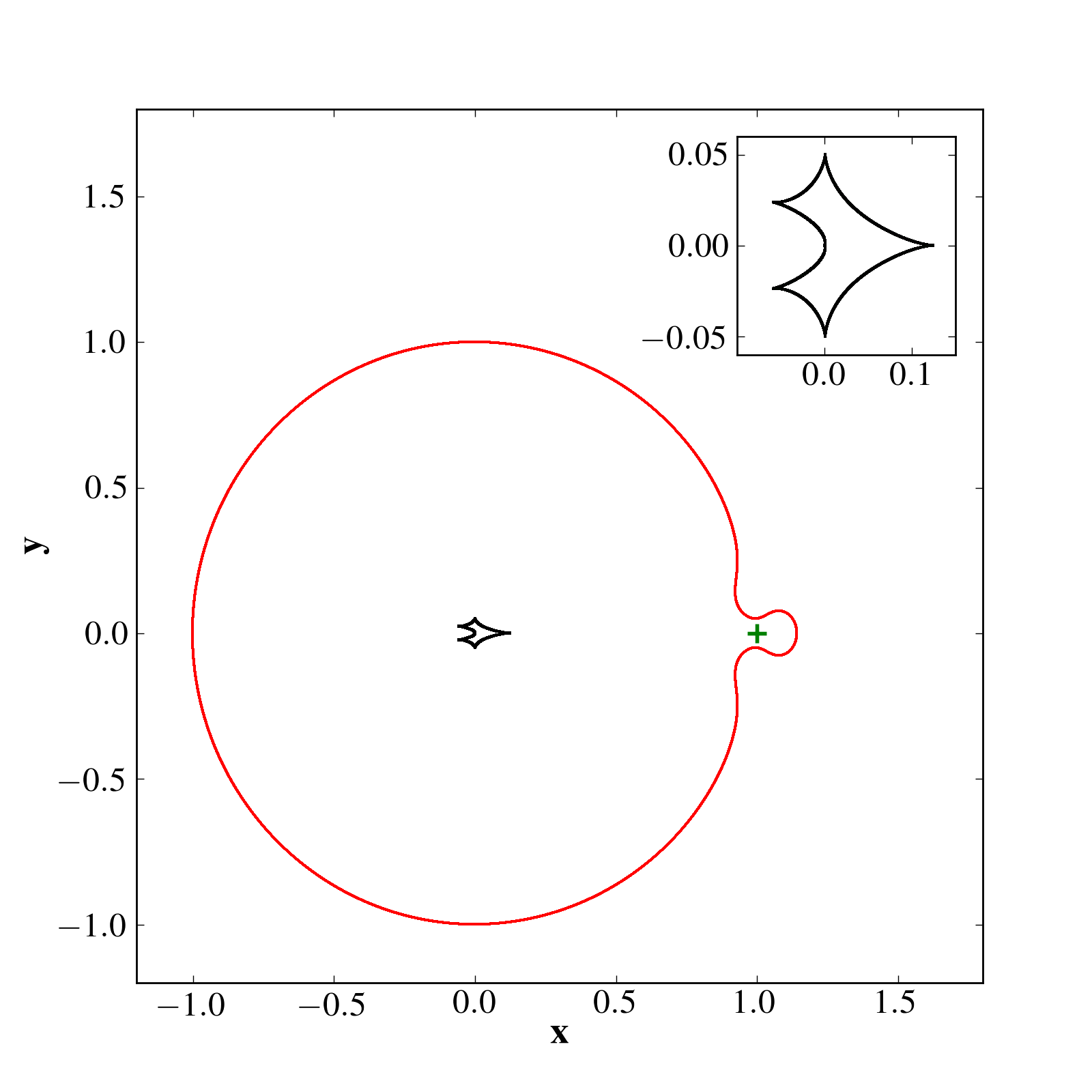}
\caption{Same as Fig. \ref{fig:critical_1_35} except $r_0=1.0$. }
\label{fig:fig_1_0}
\end{figure}

\begin{figure}
\includegraphics[scale=0.40,bb=0 0 385 567]{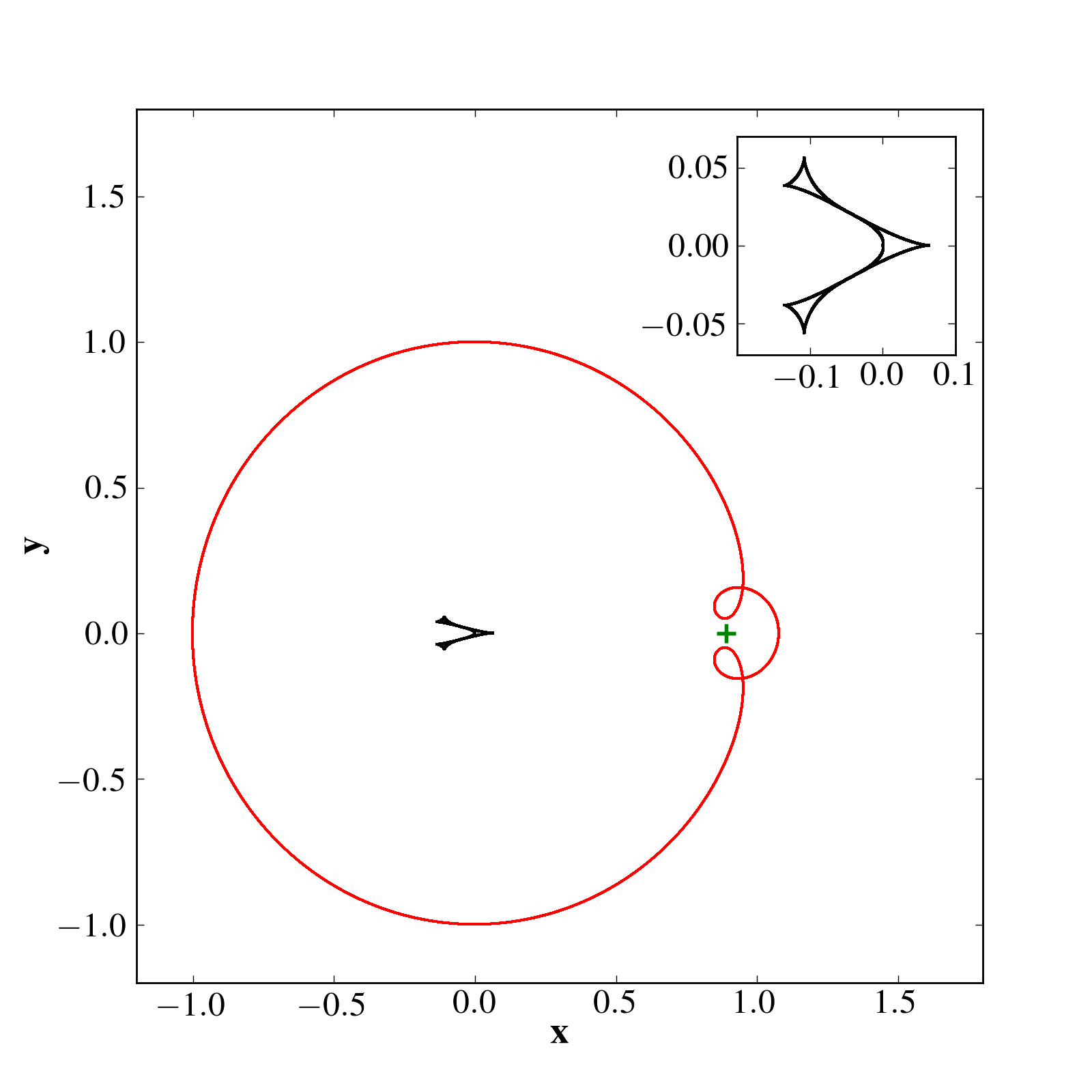}
\caption{Same as Fig. \ref{fig:critical_1_35} except $r_0=r_{t2}=0.891231$. Notice that the critical curve is just creating two ``holes" on the right, and the caustics are splitting into three segments.}
\label{fig:transition2}
\end{figure}

\begin{figure}
\includegraphics[scale=0.40,bb=0 0 385 567]{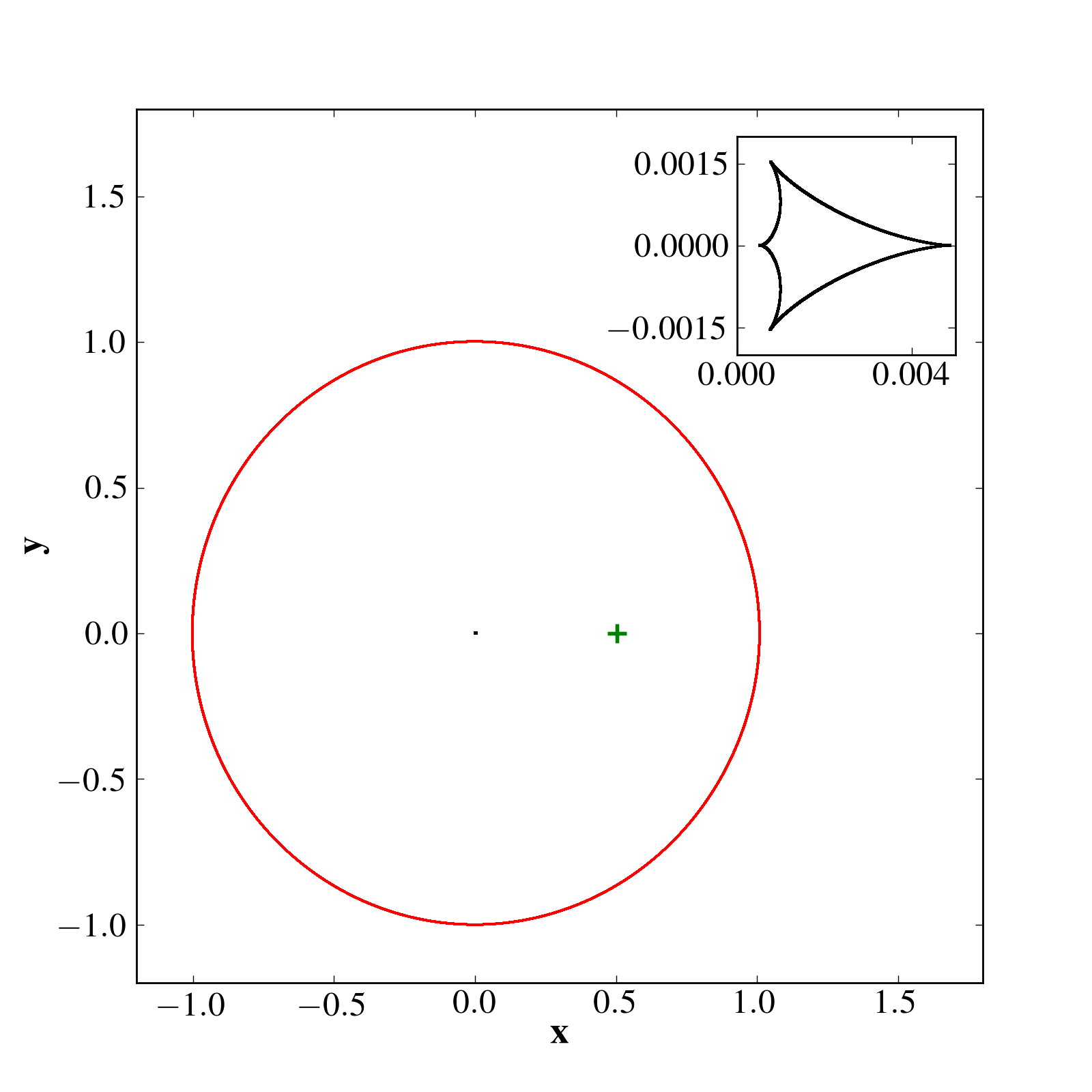}
\caption{Same as Fig. \ref{fig:critical_1_35} except $r_0=r_{t3}=0.5024$. There is only one critical curve and correspondingly one caustics.}
\label{fig:transition3}
\end{figure}

\begin{figure}
\includegraphics[scale=0.40,bb=0 0 385 567]{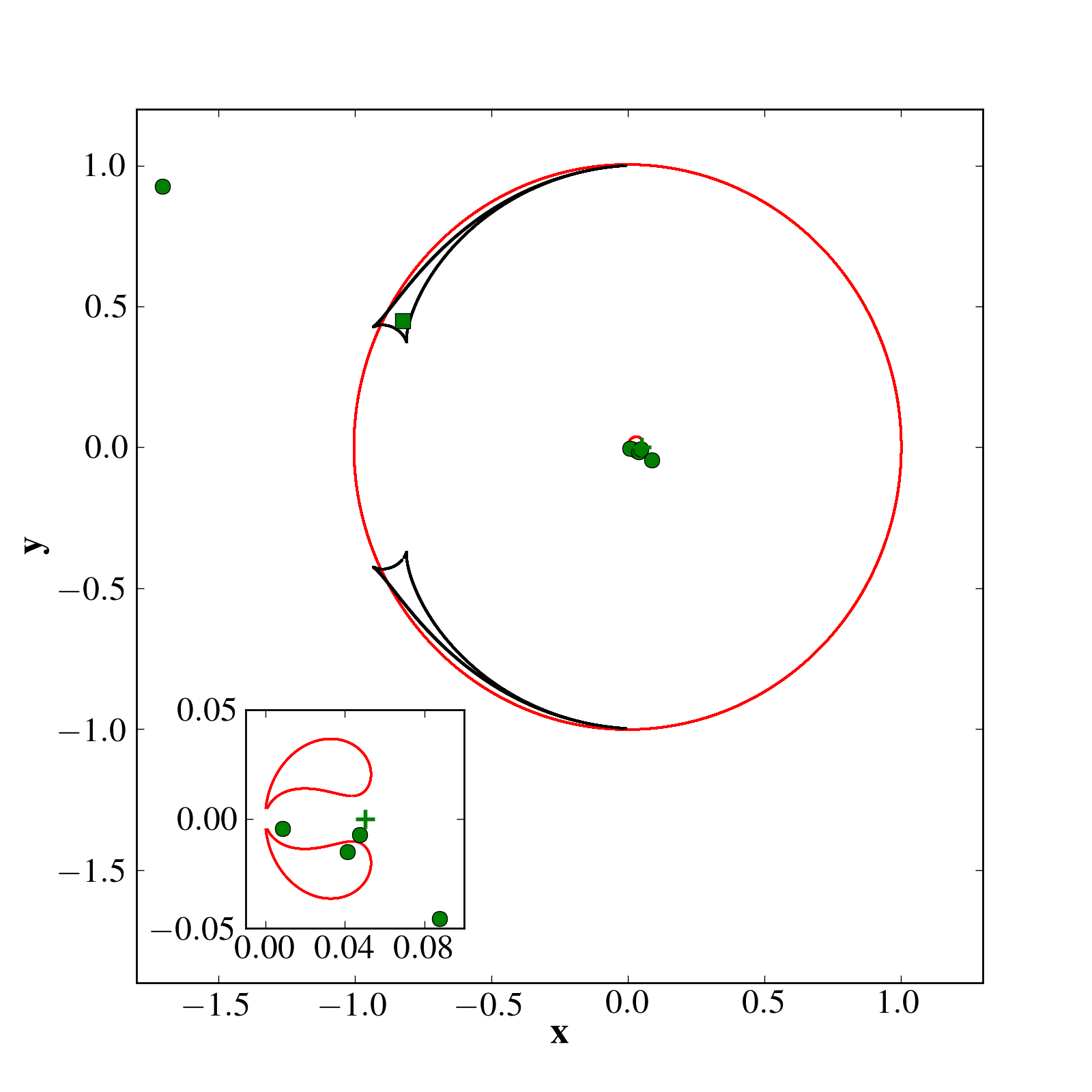}
\caption{Same as Fig. \ref{fig:critical_1_35} except $r_0=r_{t4}=0.05$. A source located at $(-0.826, 0.449)$ (green square) has five images (green circles): one positive-parity `primary' image (produced by the SIS) at the top left, and four  `central' images. The central images are better shown in the inset at the bottom left: the image close to $(0.08, -0.04)$ is the negative-parity primary image, while the other three are produced by the off-centre black hole.}
\label{fig:transition4}
\end{figure}

\begin{figure}
\includegraphics[scale=0.40,bb=0 0 385 567]{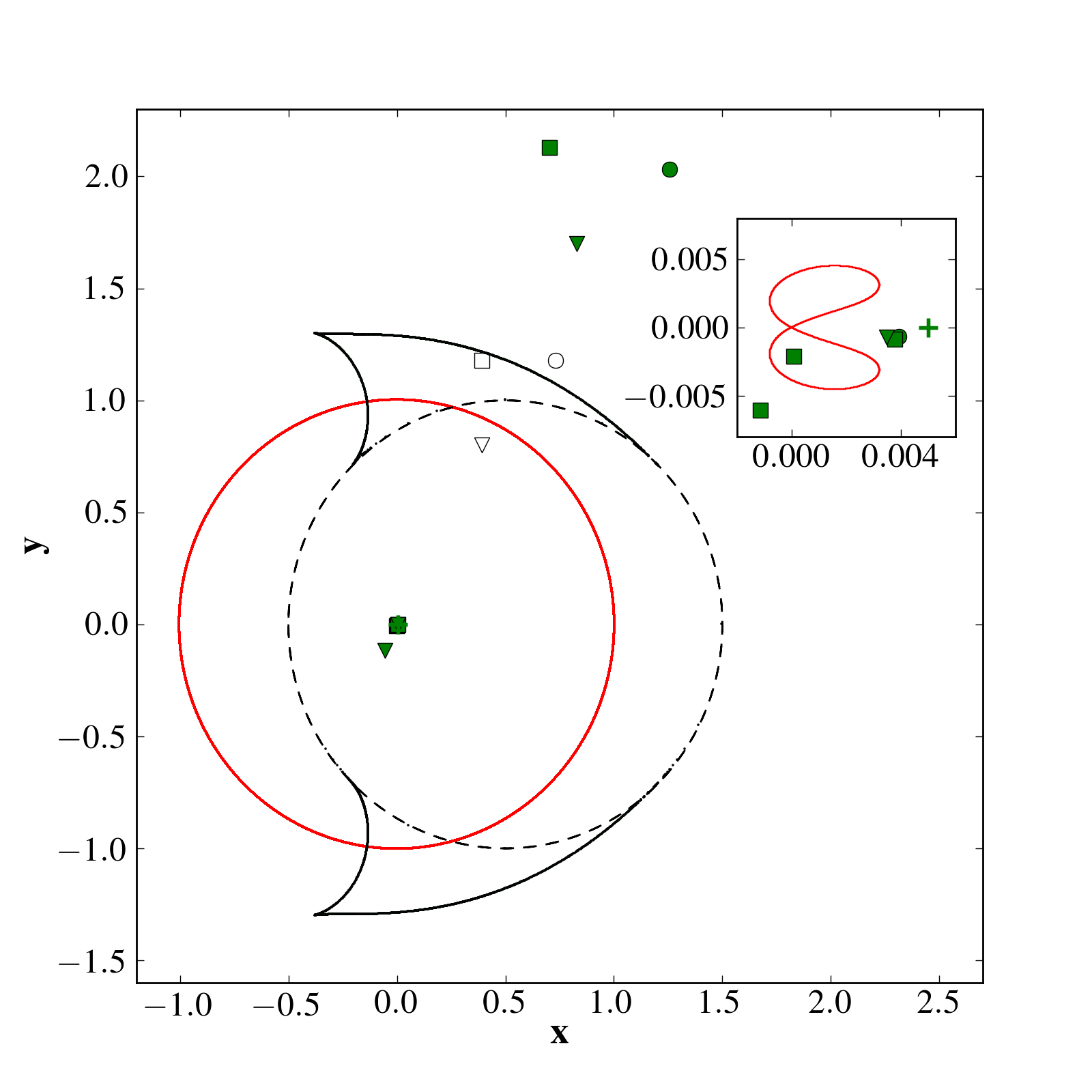}
\caption{Same as Fig. \ref{fig:critical_0_025} except $r_0=0.005$. The cross shows the point lens position. The open square, diamond and triangle symbols indicate three source positions: ($x_{\rm s}$, $y_{\rm s}$)=
(0.73 1.18), 
(0.39 1.18) and 
(0.39 0.80).
Their corresponding image positions are shown as filled symbols.  The inset shows the images close to to the central region. The red line is the critical curve. There are 4, 3, and 2 image positions for the source positions indicated by square, triangle  and diamond respectfully. The three source positions all have an image close to (0.004, 0.0). 
}
\label{fig:images}
\end{figure}

	\subsection{Examples of image configurations}
	\label{sec:images}

	We find that the image number can range from two to five. Fig. \ref{fig:images} illustrates the images for 3 source positions (open circles, triangles, and diamonds), corresponding 2, 3 and 4 images. Notice how the pseudo-caustic plays an important role. For the case labelled as a square, there are three images close to the centre. However, their magnifications are very faint, with $\mu=-3.1\times 10^{-4}$, $4.27\times 10^{-5}$ and $-8.36 \times 10^{-7}$. In comparison, the primary image  has $\mu=1.80$. The separations between them are of the order of few $\times 10^{-3}$ Einstein radius (a few milli-arcseconds) for typical galaxy lenses, which may be difficult to resolve.

	 Fig. \ref{fig:transition4} shows one five-image configuration for $r_0=r_{t4}=0.05$. 
	In this case, there are 3 faint images created by the black hole. The 
	magnifications are  $-5.98\times 10^{-3}$,  $1.16\times 10^{-2}$, $-9.50 \times 10^{-4}$ while the brightest primary image has $\mu=2.06$. Notice that the image close to $(0.08, -0.05)$ is the negative-parity image associated with the SIS.

	\section{Discussion}
	\label{sec:discussion}

	In this work, we have studied gravitational lensing by a singular isothermal sphere plus an off-centre black hole. We derived the equations for the images, critical curves and caustics. We find intriguing critical curves and caustics involving the pseudo-caustics. The total number of images for a SIS plus a single off-centre black hole can be two, three, four or five. In particular, an off-centre black hole can create a maximum of three faint images close to the centre (see Fig. \ref{fig:transition4}). To the leading order the perturbation on the magnification is quadratic on the primary images if the black hole is at the centre and linearly if it is off-centre. Our model is simplistic in modelling the primary lens galaxy as a singular isothermal sphere. While this appears to be a reasonable model for galactic-scale lenses on the scale of few kpcs (e.g., \citealt{koo09}), the central images are likely to be very sensitive to the central density profiles (\citealt{kee03, zjpb07}).  For a cored isothermal sphere, numerical investigations show that the critical curves and caustics remain similar only if the core radius (in units of the Einstein radius) is much smaller than $m$.  Magnifications are also  affected. A full investigation of an off-centre black hole in a cored isothermal sphere is beyond the scope of this paper.

	An off-centre black hole has been reported in M87 (\citealt{m87-2010}). The offset is around 12.8\,pc, of the order of $10^{-3}$ Einstein radius if we put M87 at a typical lens redshift (0.5) for a source at redshift 2. The situation will be similar to that shown in Fig. \ref{fig:images} with $r_0=0.005$. In such cases, there is a non-negligible cross-section that multiple images at the centre will be formed due to the black hole. While these images are rather faint to detect currently, they may be observable in the era of the Square Kilometer Array (SKA\footnote{www.ska.org}) where its resolution can reach milli-arcseconds and the dynamical range can be as high as a million. Since the distribution of offsets is unknown, we do not attempt a more detailed calculation of the cross-section and probabilities of seeing multiple central images due to black holes, which may also be produced by binary black holes (not yet coalescenced)
	at the centre of galaxies (\citealt{lm11}).

	\section*{Acknowledgments}
	We thank Jin An and an anonymous referee for very helpful comments and criticisms that improved the paper. We acknowledge the Chinese Academy of Sciences for financial support and the hospitalities of the Aspen Center for Physics where this work was completed.


	{\onecolumn
	\appendix
	\section{The starting point of the parametric representation}
	\label{sec:parastart}

	Using eq. (\ref{para}) and its complex conjugate for $\alpha =0$
	one can derive one equation in $r$ and another in $\cos\varphi$ using the resultant method (\citealt{es93}).
	For $r$ we obtain the following polynomial:
	\begin{equation}
	4 r^8 - 4 r^7 + (1+4m-4r_0^2) r^6 -4mr^5+ (m+r_0^2) r^4 + 4r_0^2(m+r_0^2) r^3
	- r_0^2(2m+r_0^2) r^2 + r_0^4(m-r_0^2) = 0. \label{eq:start_r}
	\end{equation} 
	For $u \equiv \cos\varphi$ we obtain
	\begin{eqnarray}
	64 r_0^4 u^8 - 32 r_0^3 u^7 - 4 r_0^2(32 r_0^2 - 4m-1)
	u^6 - 16 r_0^3 (2r_0^2 -2m -3)u^5+ 4 r_0^2 (22 r_0^2
	-11m-1) u^4+ \nonumber \\
	4 r_0 (m + 6m^2 -4r_0^2 -10 m r_1^2 +12 r_0^4) u^3 +
	(4r_0^6 -12m r_0^4 -24 r_0^4 + 12m^2 r_0^2 + 16mr_0^2 -4m^3-m^2) u^2
	\nonumber \\
	 -4r_0 (m-4r_0^2) (m-r_0^2) u-4r_0^2 (m-r_0^2)^2=0. \label{eq:start_phi}
	\end{eqnarray}
	Again one sees that when $r_0=\sqrt{m}$, $r=0$ becomes a solution of eq. (\ref{eq:start_r}), which signals one of the transitions in the topology of critical curves (see \S\ref{sec:r=0}).

	\section{Topological changes in the critical curves}
	\label{sec:appendix}

	Following \cite{es93}, the topology of critical curves changes when the following conditions are satisfied
	\begin{equation}
	J=0, 
	~~~~ \frac{\partial{J}}{\partial{\varphi}}=0, 
	~~~~\frac{\partial{J}}{\partial{r}}=0,
	\end{equation}
	where $J$ is given by eq. (\ref{eq:J}). For the derivative with respect to $\varphi$, we have
	\begin{equation}
	{\tiny
	\frac{\partial{J}}{\partial{\varphi}}=\frac{2 m r_0 \sin\varphi \,
	(2 m r^2 + r^3 - 3 r r_0^2 + 2r_0^3 u)}{r w^3}=0, ~~ u\equiv \cos\varphi, ~~ w \equiv r^2 - 2 r_0 r \cos\varphi + r_0^2.
	}
	\label{eq:Jphi}
	\end{equation}
	Similarly we find
	\begin{equation}
	J=\frac{-m^2 r + (-1 + r) w^2 - 
	 m (w -2 r_0^2 (1 - u^2))}{r w^2}=0
	 \label{eq:J2}
	\end{equation}
	and
	\begin{equation}
	\frac{\partial{J}}{\partial{r}}=\frac{4m^2 r^2 (r-r_0 u) + w^3 +m (3 r^4 - 10 r^3 r_0 u + 6 r r_0^3 u (1 - 2 u^2) + r_0^4 (-1 + 2 u^2) + 
	   6 r^2 r_0^2 (-1 + 3 u^2)) }{r^2 w^3}=0.
	   \label{Jr}
	\end{equation}
	Eq. (\ref{eq:Jphi}) is satisfied for either
	\begin{equation}
	\sin\varphi=0
	\label{eq:condition1}
	\end{equation}
	or
	\begin{equation} 
	u=-(2 m r^2 + r^3 - 3 r r_0^2)/(2r_0^3).
	\label{eq:condition2}
	\end{equation}

	For the condition in eq. (\ref{eq:condition1}), we have either $\varphi=0$ or $\varphi=\pi$. 
	For each case, the conditions $J=0$ and 
	${\partial{J}}/{\partial{r}}=0$ give two equations in terms of $r$, we can use the resultant method (\citealt{es93}) to eliminate $r$ to find the condition for topological changes in the critical curve. 
	For $\varphi=0$, $u \equiv \cos\varphi=1$, we have
	\begin{equation}
	+4 m^8 r_0^3 (m+r_0^2) (4 m + (1 - 2 r_0)^2)^2 \left[
	m + 4 m^2 - 4 r_0 - 20 m r_0 + 12 r_0^2 - 8 m r_0^2 - 12 r_0^3 + 4 r_0^4 \right]=0, 
	   \label{eq:phi0}
	   \end{equation}
	All the terms in front of the bracket [\,] are positive - only the bracket term may yield physical solutions, which can be solved analytically since it is a quartic equation in terms of $r_0$.
	Similarly, for $\varphi=\pi$ we have 
	\begin{equation}
	-4 m^8 r_0^3 (m + r_0^2) (4 m + (1+2r_0)^2)^2 \left[m + 4 m^2 + 
	   4 r_0 + 20 m r_0 + 12 r_0^2 - 8 m r_0^2 + 12 r_0^3 + 4 r_0^4\right]=0.
	\end{equation}
	For physical situations,  the black hole mass most likely satisfies $m<1$, and all terms in this equation are positive (using $12r_0^2 > 8m r_0^2$), and so there are no physical solutions.

	For the condition in eq. (\ref{eq:condition2}), the resultant method gives 
	\begin{eqnarray}
	 & 32 m^6 r_0^{10} ~ (1 + 4 m - 4 r_0^2)^2 ~ (m - r_0^2)^2  
	(8 m^4 + 91 m^5 + 344 m^6 + 
	 432 m^7 +  
	 (16 m^2 + 318 m^3 + 1659 m^4 + 2976 m^5 \nonumber \\
	   & +1296 m^6) r_0^2 +
	    (256 m + 2424 m^2 + 7333 m^3 + 7248 m^4 + 
	    864 m^5) r_0^4 + 
	    (1024 + 6208 m + 8517 m^2 - 2272 m^3 -  \nonumber \\
	 &   864 m^4) r_0^6  
	    - (3072 + 10304 m + 744 m^2 +
	    1296 m^3) r_0^8 + (3072 - 3072 m - 432 m^2) r_0^{10} - 1024 r_0^{12})=0.
	    \label{eq:phi}
	\end{eqnarray}
	This equation has at least two analytical solutions from the first two terms in brackets (given below as $r_{t3}$ and $r_{t4}$). 

	For $m=2.5\times 10^{-3}$, using eqs. (\ref{eq:phi0}-\ref{eq:phi}) we find four positive physical solutions of $r_0$
	\begin{enumerate}
	\item
	$r_{t1}=1.26361$ from the condition in eq. (\ref{eq:phi0}).
	\item $r_{t2}=0.891231$, another solution from eq. (\ref{eq:phi}) in adddition $r_{t3}$ and $r_{t4}$ below.
	\item $r_{t3}=\sqrt{1/4+m}$.
	\item $r_{t4}=\sqrt{m}$.
	\end{enumerate}
	A fifth positive solution $r_0=0.00062461$ (from eq. \ref{eq:phi0}) does not any give positive solution of $r$, and so is discarded. These four transitions are illustrated in 
	Figures \ref{fig:transition1}, \ref{fig:transition2}, \ref{fig:transition3}, and \ref{fig:transition4}. 

	\section{Maximum number of images}
	\label{sec:maximum}

	To get a maximum of six images, each equation from (\ref{eq:varphi1}) to (\ref{eq:cosPhi}) must yield two solutions. Since $r\ge 0$ and $-1 \le \cos\varphi \le 1$,  to have two solutions, eq. (\ref{eq:cosPhi}) must satisfy 
	\begin{equation}
	r_0>\rs\ge0, ~~~ m \ge \rs r_0 \frac{(1+\rs-r_0)^2}{r_0-\rs}. \label{eq:condition3}
	\end{equation}
	Furthermore for eq. (\ref{eq:varphi2}) to have two positive solutions, we must have the coefficient for the linear term to be negative
	\begin{equation}
	\rs+r_0<1 \label{eq:condition4}
	\end{equation}
	and the constant term to be positive
	\begin{equation}
	-m-r_0+\rs r_0 >0. \label{eq:condition5}
	\end{equation}
	However, combining the conditions in eqs. (\ref{eq:condition3}) and (\ref{eq:condition4}) we find that the left hand side of eq. (\ref{eq:condition5}) satisfies
	\begin{equation}
	-m-r_0+\rs r_0 \le -\rs r_0 \frac{(1+\rs- r_0)^2}{r_0-\rs} - r_0 +\rs r_0 = (-3+r_0+\rs)\rs - \frac{r_0}{r_0-\rs} < 0,
	\end{equation}
	in direct contradiction with the requirement in eq. (\ref{eq:condition5}). In other words, this equation cannot have two solutions, and so the maximum number of images is at most five, as we argued intuitively in \S\ref{sec:images}.

	\section{Magnification of the images when the source is on-axis}
	\label{sec:magOnAxis}

	When the source is located on the (positive) $x_{\rm s}$-axis ($\varphi_s =0$),
	the lens equation still factorizes (see eqs. \ref{eq:varphi1} to \ref{eq:cosPhi}); the image positions can be derived analytically and we can use eq. (\ref{eq:J}) to obtain the magnification $\mu=J^{-1}$.

	For $\varphi = 0$ we can find the solution for $r$ by solving the quadratic equation in eq. (\ref{eq:varphi1}) and then obtain the magnification using eq. (\ref{eq:J2}) ($u \equiv \cos\varphi=1$)
	\begin{equation}
	\mu=J^{-1}= \frac{r (r - r_0)^4}{-m^2 r - m (r - r_0)^2 + (-1 + r) (r - r_0)^4}. \label{eq:D1}
	\end{equation}
	Similarly for $\varphi = \pi$, we find
	\begin{equation}
	\mu = \frac{r (r + r_0)^4}{-m^2 r - m (r + r_0)^2 + (-1 + r) (r + r_0)^4}, \label{eq:D2}
	\end{equation}
	where the solutions for $r$ can be found from eq. (\ref{eq:varphi2}). For $m \ll 1$, one can Taylor expand these expressions, and find that the magnification has a linear perturbation term with respect to the black hole mass ($m$) for the two outer (primary) images:
	\begin{equation}
	\mu = \frac{\rs \pm 1}{\rs} \pm \frac{r_0}{\rs^2 (1-r_0+\rs)^2} m + O(m^2),
	\end{equation}
	where the $+$ and $-$ signs are for $\varphi=0$ and $\pi$ (eqs. \ref{eq:D1} and \ref{eq:D2}) respectively. There is one new on-axis image created by the black hole whose magnification scales as $m^2$, but we do not give the expansion here.

	The magnifications for the two off-axis images are identical due to symmetry and are given by
	\begin{equation}
	\mu = -{ 2 m r_0^3 (r_0 - r_s) \over  A_1 A_2 },
	\end{equation}
	where
	\begin{equation}
	A_{1,2} = m (r_0 - r_s)^2 - r_0 r_s (r_0 - r_s \mp 1 )^2.
	\end{equation}
	It is interesting to note that the magnification is of order $m$. These expressions are only valid when off-axis images exist, i.e., when the source is inside a tiny caustic close to the centre.

	\bibliographystyle{mn2e}
	\bibliography{./lens}

\begin{thebibliography}{}

\bibitem[\protect\citeauthoryear{{Batcheldor}, {Robinson}, {Axon}, {Perlman} \&
  {Merritt}}{{Batcheldor} et~al.}{2010}]{m87-2010}
{Batcheldor} D.,  {Robinson} A.,  {Axon} D.~J.,  {Perlman} E.~S.,    {Merritt}
  D.,  2010, \apjl, 717, L6

\bibitem[\protect\citeauthoryear{{Bowman}, {Hewitt} \& {Kiger}}{{Bowman}
  et~al.}{2004}]{2004ApJ...617...81B}
{Bowman} J.~D.,  {Hewitt} J.~N.,    {Kiger} J.~R.,  2004, \apj, 617, 81

\bibitem[\protect\citeauthoryear{{Chen}}{{Chen}}{2003a}]{2003ApJ...587L..55C}
{Chen} D.,  2003a, \apjl, 587, L55

\bibitem[\protect\citeauthoryear{{Chen}}{{Chen}}{2003b}]{2003A&A...397..415C}
{Chen} D.,  2003b, \aap, 397, 415

\bibitem[\protect\citeauthoryear{{Colpi} \& {Dotti}}{{Colpi} \&
  {Dotti}}{2009}]{2009arXiv0906.4339C}
{Colpi} M.,  {Dotti} M.,  2009, ArXiv e-prints

\bibitem[\protect\citeauthoryear{{Erdl} \& {Schneider}}{{Erdl} \&
  {Schneider}}{1993}]{es93}
{Erdl} H.,  {Schneider} P.,  1993, \aap, 268, 453

\bibitem[\protect\citeauthoryear{{Evans} \& {Wilkinson}}{{Evans} \&
  {Wilkinson}}{1998}]{ew98}
{Evans} N.~W.,  {Wilkinson} M.~I.,  1998, \mnras, 296, 800

\bibitem[\protect\citeauthoryear{{Gualandris} \& {Merritt}}{{Gualandris} \&
  {Merritt}}{2008}]{gua08}
{Gualandris} A.,  {Merritt} D.,  2008, \apj, 678, 780

\bibitem[\protect\citeauthoryear{{G{\"u}ltekin}, {Richstone}, {Gebhardt},
  {Lauer}, {Tremaine}, {Aller}, {Bender}, {Dressler}, {Faber}, {Filippenko},
  {Green}, {Ho}, {Kormendy}, {Magorrian}, {Pinkney} \& {Siopis}}{{G{\"u}ltekin}
  et~al.}{2009}]{bh2009}
{G{\"u}ltekin} K.,  {Richstone} D.~O.,  {Gebhardt} K.,  {Lauer} T.~R.,
  {Tremaine} S.,  {Aller} M.~C.,  {Bender} R.,  {Dressler} A.,  {Faber} S.~M.,
  {Filippenko} A.~V.,  {Green} R.,  {Ho} L.~C.,  {Kormendy} J.,  {Magorrian}
  J.,  {Pinkney} J.,    {Siopis} C.,  2009, \apj, 698, 198

\bibitem[\protect\citeauthoryear{{Keeton}}{{Keeton}}{2003}]{kee03}
{Keeton} C.~R.,  2003, \apj, 582, 17

\bibitem[\protect\citeauthoryear{{Koopmans}, {Bolton}, {Treu}, {Czoske},
  {Auger}, {Barnab{\`e}}, {Vegetti}, {Gavazzi}, {Moustakas} \&
  {Burles}}{{Koopmans} et~al.}{2009}]{koo09}
{Koopmans} L.~V.~E.,  {Bolton} A.,  {Treu} T.,  {Czoske} O.,  {Auger} M.~W.,
  {Barnab{\`e}} M.,  {Vegetti} S.,  {Gavazzi} R.,  {Moustakas} L.~A.,
  {Burles} S.,  2009, \apjl, 703, L51

\bibitem[\protect\citeauthoryear{{Li}, {Mao}, {Gao}, {Loeb} \& {Di
  Stefano}}{{Li} et~al.}{2011}]{lm11}
{Li} N.,  {Mao} S.,  {Gao} L.,  {Loeb} A.,    {Di Stefano} R.,  2011, \mnras,
  submitted

\bibitem[\protect\citeauthoryear{{Mao}, {Witt} \& {Koopmans}}{{Mao}
  et~al.}{2001}]{mwk01}
{Mao} S.,  {Witt} H.~J.,    {Koopmans} L.~V.~E.,  2001, \mnras, 323, 301

\bibitem[\protect\citeauthoryear{{Merritt} \& {Milosavljevi{\'c}}}{{Merritt} \&
  {Milosavljevi{\'c}}}{2005}]{2005LRR.....8....8M}
{Merritt} D.,  {Milosavljevi{\'c}} M.,  2005, Living Reviews in Relativity, 8,
  8

\bibitem[\protect\citeauthoryear{{Pretorius}}{{Pretorius}}{2007}]{pre07}
{Pretorius} F.,  2007, ArXiv e-prints

\bibitem[\protect\citeauthoryear{{Rusin}, {Keeton} \& {Winn}}{{Rusin}
  et~al.}{2005}]{2005ApJ...627L..93R}
{Rusin} D.,  {Keeton} C.~R.,    {Winn} J.~N.,  2005, \apjl, 627, L93

\bibitem[\protect\citeauthoryear{{Tsygan}}{{Tsygan}}{2007}]{tsy07}
{Tsygan} A.~I.,  2007, Astronomy Reports, 51, 97

\bibitem[\protect\citeauthoryear{{Witt}}{{Witt}}{1990}]{wit90}
{Witt} H.~J.,  1990, \aap, 236, 311

\bibitem[\protect\citeauthoryear{{Yu}}{{Yu}}{2002}]{2002MNRAS.331..935Y}
{Yu} Q.,  2002, \mnras, 331, 935

\bibitem[\protect\citeauthoryear{{Zhang}, {Jackson}, {Porcas} \&
  {Browne}}{{Zhang} et~al.}{2007}]{zjpb07}
{Zhang} M.,  {Jackson} N.,  {Porcas} R.~W.,    {Browne} I.~W.~A.,  2007,
  \mnras, 377, 1623

\end{thebibliography}

	\label{lastpage}
\end{document}